\documentclass[pra]{revtex4}
\usepackage{amsmath}

\newcommand{\vc}[1]{\boldsymbol{#1}} 

\begin{document}

\title{Derivation of the magnetization current from the non-relativistic
Pauli equation: A comment on ``The quantum mechanical current of
the Pauli equation'' by Marek Nowakowski [Am.~J.~Phys.~67(10),
916-919 (1999)] }
\author{M. S. Shikakhwa}
\affiliation{ Physics Program, Middle East Technical University Northern Cyprus Campus,\\
Kalkanl\i, G\"{u}zelyurt, TRNC, via Mersin 10, Turkey}
\author{S. Turgut}
\author{N. K. Pak}
\affiliation{Department of Physics, Middle East Technical University,\\
TR-06800, Ankara, Turkey}

\begin{abstract}

\end{abstract}



\maketitle

Some time ago, Nowakowski \cite{Nowakowski} presented a discussion
of the fact that in the non-relativistic limit, the probability
current density $\vc{J}$ of a spin one-half particle contains an
extra divergenceless term $\vc{J}_M$
\begin{eqnarray}
  \vc{J}&=& \vc{J}_0 + \vc{J}_M  \nonumber \\
        &=&  \frac{\hbar}{2mi}\left(\psi^\dagger  \vc{\nabla}\psi-(\vc{\nabla}\psi^\dagger)\psi\right)
            +  \frac{\hbar}{2m}\vc{\nabla}\times(\psi^\dagger\vc{\sigma}\psi)\quad.
  \label{eq:J_expr}
\end{eqnarray}
which he derived by taking the non-relativistic limit of the
relativistic Dirac probability current density. The derivation for
this additional term essentially relies on two assumptions: (1) In
Dirac equation, the probability density of the particle is given
by $\rho=\psi^\dagger\psi$ and (2) $\rho$ is the time-component of
a Lorentz covariant four-vector $(\rho,\vc{J})$. Therefore, these
assumptions uniquely identify the probability current density.
Taking the non-relativistic limit of this current produces
Eq.~(\ref{eq:J_expr}).

Simply because of its \emph{existence}, $\vc{J}_M$ term is
important and hence it should be included in textbook discussions
of the probability current of spin 1/2 particles. A nice
discussion of such an additional current term with illustrative
examples from various quantum mechanical systems was published in
this journal\cite{Mita} (But, note that this work defines the
additional term from a different perspective and thus differs by a
factor of 2 from the correct $\vc{J}_M$).

Nowakowski correctly states that the additional term $\vc{J}_M$
cannot be derived from the non-relativistic Pauli equation, as the
covariance argument can only be applied at the fully-relativistic
Dirac equation level. Even though this is the correct state of the
affairs, it is still desirable to have an alternative derivation
of this additional current term from a non-relativistic ``starting
point''. If one is to derive the $\vc{J}_M$ term for undergraduate
or junior graduate students who have not yet been exposed to
relativistic quantum mechanics, one needs to start from the Pauli
equation. The purpose of this comment is to point out that there
is indeed an alternative derivation of this additional term
starting from the non-relativistic quantum mechanics of a spin 1/2
particle.

Our starting point is an alternative form of the Pauli
Hamiltonian, namely
\begin{equation}
    H =\frac{1}{2m}(\vc{\sigma}\cdot\vc{p})^2
    \label{eq:altPauli}
\end{equation}
where $\vc{p}=-i\hbar\vc{\nabla}$ is the momentum operator and $\sigma_i$
($i=1,2,3$) are the Pauli spin matrices. Using the well-known identity
\begin{equation}
    (\vc{\sigma}\cdot\vc{u})(\vc{\sigma}\cdot\vc{v})=\vc{u}\cdot\vc{v}+i\vc{\sigma}\cdot(\vc{u}\times\vc{v})\quad,
    \label{eq:SigmaIdentity}
\end{equation}
which can be easily derived from 
\begin{equation}
  \sigma_i\sigma_j=\delta_{i,j}I+i\sum_k\epsilon_{ijk}\sigma_k~~,
  \label{eq:SigmaIdentityIndicial}
\end{equation}
it can be seen that Eq.~(\ref{eq:altPauli}) is the same as the
Hamiltonian $H=\vc{p}^2/2m$. The form in Eq.~(\ref{eq:altPauli}),
however, has the obvious advantage that, for a charged particle,
in the presence of coupling to a vector potential $\vc{A}$ (so
that $\vc{p}$ is replaced with $\vc{\pi}=\vc{p}-(e/c)\vc{A}$), we
have
\begin{eqnarray}
 H &=& \frac{1}{2m}(\vc{\sigma}\cdot\vc{\pi})^2 \\
    &=& \frac{1}{2m}\vc{\pi}^2 -\frac{e\hbar}{2mc}\vc{\sigma}\cdot\vc{B}\quad,
\end{eqnarray}
i.e., the Zeeman term in the Hamiltonian is generated
automatically with the correct g-factor of $g=2$ (see, for example
Ref.~\onlinecite{Sakurai}), rather that being introduced by hand
as a phenomenological term, as is usually done. This is intimately
related with the fact that Eq.~(\ref{eq:altPauli}) is the first
expression obtained for the Hamiltonian when the non-relativistic
limit of the Dirac equation is taken, before simplifying it
further into the original Pauli Hamiltonian. This teaches us that 
 Eq.~(\ref{eq:altPauli}) is the fundamental non-relativistic 
Hamiltonian that one should start with; the form $H=p^2/2m$ is just a reduced special case. 
Coupled with these,
if we start from the Hamiltonian in Eq.~(\ref{eq:altPauli}), and
if we are careful in not canceling some terms, it is possible to
derive the additional term $\vc{J}_M$ in the probability current
density. Below, we present an alternative derivation of the
current density based on the conventional continuity equation. For
the sake of completeness, we also sketch a second derivation based
on Noether's theorem.

As the probability current density is usually derived from the
continuity equation, it is important to show that this approach
also produces $\vc{J}_M$. Note that, since
$\vc{\nabla}\cdot\vc{J}_M=0$, this term obviously does not have
any contribution to the continuity equation,
$\partial(\psi^\dagger\psi)/\partial t +\vc{\nabla}\cdot\vc{J}=0$.
For this reason, one has to be careful in not dropping some
relevant terms. It is important to remember Nowakowski again: in a
non-relativistic derivation, one cannot understand the presence,
form or the coefficient of such additional terms. However, we will
show that ``the additional term is already there'' before it is
swept away (canceled) under the divergence operator.

In this approach, it is very useful to consider the case where
there is a vector potential $\vc{A}$ so that the momentum $\vc{p}$
are replaced with $\vc{\pi}$. Of course, for chargeless particles
(e.g., neutrinos), such a change is not physically allowed. Our
main purpose for incorporating such a change is to remind us not
to commute different components of $\vc{\pi}$. By keeping in mind
that they do not commute, the operators will naturally guide us
through the derivation. It will be seen that the vector potential
will never be important in any stage of the derivation. At the end
of the derivation, the vector potential can be set to zero if
necessary. This is a reflection of the fact that the derivation is also valid
for chargeless particles. Also, the Hamiltonian on the right-hand side of
Eq.~(\ref{eq:altPauli}) may contain a scalar potential term. As it
has no effect on the current density, we omit such a potential
term in the following derivation. This derivation starts with the
conventional construction of the continuity equation, namely by
taking the time derivative of the probability density
$\rho=\psi^\dagger\psi$,
\begin{eqnarray}
  \frac{\partial\rho}{\partial t}  &=&  \frac{\partial\psi^\dagger}{\partial t}\psi
                    +   \psi^\dagger \frac{\partial\psi}{\partial t} \\
      &=& -\frac{1}{2mi\hbar}\Big( ((\vc{\sigma}\cdot\vc{\pi})^2\psi)^\dagger \psi - \psi^\dagger(\vc{\sigma}\cdot\vc{\pi})^2\psi \Big) \\
      &=& -\frac{1}{2mi\hbar}\sum_{i,j}\Big( (\pi_i\pi_j\psi)^\dagger \sigma_j\sigma_i\psi - \psi^\dagger\sigma_i\sigma_j (\pi_i\pi_j\psi) \Big)
\end{eqnarray}
It is important to note that the non-commutativity of the kinetic
momenta $\pi_i$ have guided us to express the two terms as above.
Otherwise, if the vector potential were set to zero at the
beginning, we would have no rational reason for preferring $p_i
p_j\psi$ over $p_jp_i\psi$. We will now try to express the
right-hand side as the divergence of a current by ``pulling up''
the derivative $\partial/\partial x_i$ to the front.
The following identity is very useful for this purpose and, aside
from factors involving the vector potential, is obtained from the
chain rule of differentiation
\begin{equation}
  \alpha^\dagger (\pi_i\beta) - (\pi_i\alpha)^\dagger \beta
        =\frac{\hbar}{i}\frac{\partial}{\partial x_i}\left(\alpha^\dagger \beta\right)\quad.
  \label{eq:Valuable_identity}
\end{equation}
In the above equation $\alpha$ and $\beta$ are arbitrary
two-component spinors. Note that the vector potential does not
occur on the right-hand side of the identity. (It is useful to
think that this identity follows from the fact that $\pi_i$ is
hermitian: as the integral of the left-hand side must be zero, the
right-hand side must be a divergence.) Using this we get
\begin{eqnarray}
  \frac{\partial\rho}{\partial t} &=&  -\frac{1}{2m}\sum_{i,j} \frac{\partial}{\partial x_i}
       \Big( (\pi_j\psi)^\dagger \sigma_j\sigma_i\psi + \psi^\dagger\sigma_i\sigma_j (\pi_j\psi)  \Big)
       \nonumber  \\
       & &\quad
         - \frac{1}{2mi\hbar}\sum_{i,j}
           \Big( (\pi_j\psi)^\dagger \sigma_j\sigma_i(\pi_i\psi) - (\pi_i\psi)^\dagger\sigma_i\sigma_j (\pi_j\psi)  \Big)
         \label{eq:Stop_Here}
\end{eqnarray}
Now, the second sum in the equation above gives zero (this can be
seen easily by exchanging the labels $i\leftrightarrow j$ for one
of the summands). The first sum is in the desired divergence form
and the probability-current vector can be read directly as
\begin{equation}
  J_i = \frac{1}{2m}\sum_j (\pi_j\psi)^\dagger \sigma_j\sigma_i\psi + \psi^\dagger\sigma_i\sigma_j  (\pi_j\psi)\quad;
  \label{eq:Current_Density}
\end{equation}
the continuity equation $\partial\rho/\partial t
+\vc{\nabla}\cdot\vc{J}=0$ is then satisfied. What is left is the
simplification of the right-hand side to the sum of the
conventional and the magnetization current. Using
Eq.~(\ref{eq:SigmaIdentityIndicial}) we
get
\begin{eqnarray}
  J_i  &=&  \frac{1}{2m} \Big((\pi_i\psi)^\dagger \psi +  \psi^\dagger(\pi_i\psi)\Big)
           + \frac{i}{2m}\sum_{j,k}\epsilon_{ijk} \Big(-(\pi_j\psi)^\dagger \sigma_k\psi + \psi^\dagger\sigma_k(\pi_j\psi)\Big) \\
   &=& J_{0i}+\frac{i}{2m}\sum_{j,k}\epsilon_{ijk} \frac{\hbar}{i} \frac{\partial}{\partial x_j} \Big( \psi^\dagger \sigma_k \psi\Big) \\
   &=& J_{0i} + \frac{\hbar}{2m}\left(\vc{\nabla}\times\psi^\dagger\vc{\sigma}\psi \right)_i=J_{0i}+J_{Mi}\quad.
\end{eqnarray}
Here, we have used the identity in
Eq.~(\ref{eq:Valuable_identity}) again for simplifying the
magnetization current term. Note that the correct expression of
$\vc{J}_0$ in the presence of a vector potential is as above
(i.e., $\vc{J}$ is the real part of $\psi^\dagger\vc{\pi}\psi/m$)
which basically reduces to the familiar form when $\vc{A}$ is set
to zero. Note also that, at no point in the derivation above the
vector potential $\vc{A}$ appears explicitly. It is only
implicitly present by reminding us that we should not commute
products $\pi_i\pi_j$.

Note that we have not simplified the products of Pauli matrices
until the last point. At Eq.~(\ref{eq:Stop_Here}), when we have
seen that the last sum is zero, we have stopped and identified the
current density as (\ref{eq:Current_Density}). It is still
possible to not stop at Eq.~(\ref{eq:Stop_Here}), continue
simplifying the first sum in this equation and drop the vanishing
terms corresponding to $\vc{\nabla}\cdot\vc{J}_M=0$. In such a
path, one can obtain unsurprisingly only the $\vc{J}_0$ term of
the current. Of course, these kinds of ambiguities are expected as
we are not following the only correct methodology (i.e., using
covariance argument in Dirac equation). Despite this, the very
fact that one can find the correct term $\vc{J}_M$ by \emph{not
canceling} an obviously zero term in the continuity equation,
gives some credence to the current approach.


The additional current term can also be shown to be a part of the
conserved Noether current\cite{Peskin} that follows from the
invariance of the non-relativistic Pauli Lagrangian under the U(1)
global phase transformation. In this case, the presence or the
absence of the vector potential does not change the derivation.
For this reason we take $\vc{A}=0$, since a non-zero vector
potential does not change the derivation.

The Lagrangian density is given by
\begin{equation}
  \mathcal{L}=\frac{i\hbar}{2}\left(\psi^\dagger \dot{\psi}-\dot{\psi^\dagger} \psi\right)
           -\frac{1}{2m}(\vc{\sigma}\cdot\vc{p}\psi)^\dagger
                        (\vc{\sigma}\cdot\vc{p}\psi)\quad,
\end{equation}
with the Euler-Lagrange equations giving Eq.(\ref{eq:altPauli}). Note that, upon simplification of the spin
matrices, the Lagrangian density can be written as
$\mathcal{L}=\mathcal{L}_0+\mathcal{L}_M$ where
\begin{eqnarray}
  \mathcal{L}_0  &=&   \frac{i\hbar}{2}\left(\psi^\dagger \dot{\psi}-\dot{\psi^\dagger} \psi\right)
           -\frac{\hbar^2}{2m}\sum_i \partial_i \psi^\dagger           \partial_i\psi  \quad, \\
  \mathcal{L}_M         &=& -\frac{i\hbar^2}{2m}\sum_{ijk} \epsilon_{ijk}\partial_i\psi^\dagger \sigma_k \partial_j\psi\quad.
\end{eqnarray}
The last term, $\mathcal{L}_M$, can be brought into the form of a total
divergence
\begin{equation}
  \mathcal{L}_M = \frac{i\hbar^2}{2m} \vc{\nabla}\cdot(\psi^\dagger\vc{\sigma}\times\vc{\nabla}\psi)\quad,
\end{equation}
and hence its contribution to the action $I=\int\mathcal{L}d^4x$
can be converted into a surface integral. Because of this reason,
its presence in the Lagrangian density does not affect the
equations of motion. However, if this term is kept, one derives
the missing magnetization current term here too. (Note that, if
there is magnetic field, then $\mathcal{L}_M$ is not equal to a
divergence. In fact, $\mathcal{L}_M$ contains the Zeeman term in
that case.)

Now, consider the global phase transformation
\begin{equation}
  \psi \longrightarrow e^{-i\alpha/\hbar}\psi \approx
  \left(1-\frac{i}{\hbar}\delta\alpha\right)\psi
\end{equation}
where $\delta\alpha$ is an infinitesimal real number independent
of time and space. The Lagrangian density $\mathcal{L}$ is
invariant under this transformation. The corresponding conserved
Noether current is given by
\begin{equation}
  J^\mu \delta\alpha =\delta\psi^\dagger
  \frac{\partial\mathcal{L}}{\partial(\partial_\mu\psi^\dagger)}+\frac{\partial\mathcal{L}}{\partial(\partial_\mu\psi)}\delta\psi
  \quad (\mu=0,1,2,3)~.
\end{equation}
It is then straightforward to see that the time component
($\mu=0$), $J^0=\psi^\dagger\psi$, is the probability density
$\rho$ and the space components produce
$\vc{J}=\vc{J}_0+\vc{J}_M$. The additional term $\vc{J}_M$ comes
from the term $\mathcal{L}_M$. 


Taking this opportunity, we would like to comment on some points
related to the interpretation of the current $\vc{J}_M$. It is
possible to interpret this term as the effective current density
associated with magnetization in classical
electrodynamics\cite{Jackson}. In other words, if the particle is
a charged particle with charge $e$, then its magnetization density
is given by $\vc{M}=(e\hbar/2mc)\psi^\dagger\vc{\sigma}\psi$ and
the associated magnetization current is
$c\vc{\nabla}\times\vc{M}=e\vc{J}_M$. In fact, Landau and Lifshitz
have derived this term by taking the functional derivative of the
(average) energy with respect to the vector
potential\cite{LandauLifshitz}. This approach essentially
identifies the current density $\vc{J}$ as the ``coupling
strength'' of the particle to the electromagnetic field. As the
internal magnetic moment associated with the spin creates a
magnetic field, it is necessary that a spin term is also present
in $\vc{J}$.

Although the argument above for the interpretation of the
additional $\vc{J}_M$ term seems consistent, it is definitely
incomplete, as the same term is also present for chargeless spin
1/2 particle which do not couple to the electromagnetic
field, (e.g., neutrinos) as well. For chargeless particles, the
additional term is still present. This also explains ``the
coincidence'' that the additive term looks like a magnetization
current: For charged particles, the charge density and the
probability density are directly proportional to each other. By
relativistic covariance, the associated currents should also be
directly proportional to each other. Therefore, if the charge
current contains a term related to the magnetization, then, there
should also be such a term in the probability current. Yet, the
origin of the additive term is not electromagnetism as it is also
present for particles that do not interact electromagnetically.

\end{document}